%% file: main.tex
\documentclass{article}

\usepackage{arxiv} 

\usepackage{siunitx} 
\usepackage{graphicx} 
\usepackage{epstopdf}

\usepackage{amsmath} 
\usepackage[utf8]{inputenc}
\usepackage[english]{babel}
\usepackage{comment}

\usepackage[sort&compress,numbers]{natbib}
  
\newcommand{\vol}{{\ooalign{\hfil$V$\hfil\cr\kern0.08em--\hfil\cr}}}

\renewcommand{\headeright}{ }

\usepackage[tableposition=top]{caption}
\usepackage{subcaption}
\usepackage{float}
\floatstyle{plaintop}
\restylefloat{table}

\usepackage[utf8]{inputenc}
\usepackage[english]{babel}
\usepackage{hyperref}
\hypersetup{
    colorlinks=true,
    linkcolor=red, 
		anchorcolor=black,
		citecolor=green,
    filecolor=cyan,
		runcolor=cyan,
		menucolor=red,
		urlcolor=magenta, 
}
\usepackage{authblk}

\usepackage{lmodern} 
\usepackage{xcolor} 

\usepackage{soul}
\definecolor{light-gray-x}{gray}{0.9}
\definecolor{pink-x}{rgb}{0.858, 0.188, 0.478}
\definecolor{blue-x}{RGB}{24, 90, 188}
\definecolor{gray-x}{RGB}{241, 243, 244}

\usepackage[export]{adjustbox}

\usepackage{listings}
\definecolor{codegreen}{rgb}{0,0.6,0}
\definecolor{codegray}{rgb}{0.5,0.5,0.5}
\definecolor{codepurple}{rgb}{0.58,0,0.82}
\definecolor{backcolour}{rgb}{0.95,0.95,0.92}
\lstdefinestyle{mystyle}{
    backgroundcolor=\color{backcolour},   
    commentstyle=\color{codegreen},
    keywordstyle=\color{magenta},
    numberstyle=\tiny\color{codegray},
    stringstyle=\color{codepurple},
    basicstyle=\ttfamily\footnotesize,
    breakatwhitespace=false,         
    breaklines=true,                 
    captionpos=b,                    
    keepspaces=true,                 
    numbers=left,                    
    numbersep=5pt,                  
    showspaces=false,                
    showstringspaces=false,
    showtabs=false,                  
    tabsize=2
}
\usepackage[ruled,linesnumbered]{algorithm2e}

\SetCommentSty{mycommfont}

\usepackage{textcomp}

\title{A tracking algorithm for finite-size particles}

\author[$\dagger$$\ast$]{Aryan Mehboudi} 
\author[$\dagger$]{Shrawan Singhal} 
\author[$\dagger$$\mathsection$]{S.V. Sreenivasan}

\affil[$\dagger$]{NASCENT Engineering Research Center,
The University of Texas at Austin, Austin, Texas 78758, United States}
\affil[$\mathsection$]{
Walker Department of Mechanical Engineering, The University of Texas at Austin, Austin, TX 78712, United States}
\affil[$\ast$]{Corresponding author \textemdash E-mail: aryan.mehboudi@austin.utexas.edu}

\date{December 18, 2024}

\begin{document}
\clearpage
\maketitle 
\thispagestyle{empty}

\input{abstract}
\newpage
\input{introduction}

\input{method}

\input{results}

\input{conclusion}

\section*{Data Availability Statement}
The developed code is accessible from 
\href{https://github.com/am-0code1/fspt}{https://github.com/am-0code1/fspt}.
The example files and manual can be found at 
\href{https://utexas.box.com/v/am-fspt-docs}{https://utexas.box.com/v/am-fspt-docs}.

\section*{Acknowledgments}
A.M. greatly acknowledges the Microelectronics Research Center (MRC) at The University of Texas at Austin
and all technical staff for their contributions to the fabrication work. 
Fluorescence microscopy was performed at the Center for Biomedical Research Support Microscopy and Imaging Facility at UT Austin (RRID:SCR\_021756) by using a Nikon SMZ25 fluorescence stereoscope.
A.M. greatly appreciates The CBRS Microscopy and Imaging Facility and all staff for their contribution to flow visualization work with special thanks to Anna Webb and Paul Oliphint for training and assistance with fluorescence microscopy.

\section*{Conflict of Interest}
The authors have no conflicts to disclose.



\input{main.bbl}



\end{document}

%% file: abstract.tex
\begin{abstract}
Particle-wall interactions play a crucially important role in various applications such as 
microfluidic devices for cell sorting, 
particle separation,
entire class of hydrodynamic filtration and its derivatives, 
etc.
Yet, accurate implementation of interactions between wall and finite-size particle is not trivial when working with the currently available particle tracking algorithms/packages as they typically work with point-wise particles.
Herein, we report a particle tracking algorithm that takes into account interactions between particles of finite size and solid objects existing inside computational domain.
A particle is modeled as a set of circumferential points on its perimeter.
While fluid-particle interactions are captured during the track of particle center, interactions between particle and nearby solid objects are modeled explicitly by examining circumferential points and 
applying a reflection scheme as needed 
to ensure impenetrability of solid objects.
We also report a modified variant of auxiliary structured grid method to locate hosting cells, which in conjunction with a boundary condition scheme enables the capture of interactions between particle and solid objects.
As a proof-of-concept, we numerically and experimentally study the motion of particles within a microfluidic deterministic lateral displacement device. 
The modeling results successfully demonstrate the zig-zag and bumping displacement modes observed in our experiments.
We also study a microfluidic device with pinched flow numerically and validate our results against experimental data from the literature.
By demonstrating an almost 8x speedup on a system with 8 Performance threads, our investigations suggest that the particle tracking algorithm and its implementation code can benefit from parallel processing on multi-thread systems by using the OpenMP application programming interface.
We believe that the proposed framework can pave the way for designing 
related microfluidic chips precisely and conveniently.
\\
\\
\noindent\textit{Keywords:} 
Particle tracking,
Microfluidics,
Hydrodynamic filtration,
Pinched flow,
Deterministic lateral displacement,
DLD,
Particle separation,
Parallel processing,
OpenMP
\end{abstract}

%% file: introduction.tex
\section{Introduction}
\label{sec_introduction}

Particle tracking is 
an essential tool used
in various fields such as studying 
transport of environmental pollutants in water and air \cite{korotenko_particle_2004, pilechi_numerical_2022},
particle transport and deposition in ventilation systems 
\cite{zhang_experimental_2006,
widiatmojo_predicting_2016,
crawford_modeling_2021,
tan_current_2022,
chu_transport_2022},
biological respiratory system 
\cite{tsuda_particle_2013, 
inthavong_lagrangian_2016, 
islam_euler-lagrange_2019, 
atzeni_computational_2021, 
mortazavy_beni_experimental_2021, 
kiasadegh_transient_2020}, 
drug delivery applications 
\cite{schuster_particle_2015,
rukshin_modeling_2017,
yuan_effect_2022,
buchete_multiscale_2024}, 
microfluidic systems for cell sorting \cite{wu_numerical_2023,timary_shear-enhanced_2023,lv_diseased_2024} 
etc.
There are typically two major strategies to treat particles when formulating numerical simulations: Lagrangian and Eulerian. 
In a Lagrangian approach, particles are considered as a discrete phase and their trajectory is tracked individually. 
The Eulerian framework, on the other hand, considers continuous phase(s) for particles together with conservation equations to formulate transport of particles concentration.
An Eulerian technique is typically less expensive and directly deals with particle concentration, which is a bulk parameter and suitable for many engineering applications \cite{saidi_comparison_2014, xu_comparison_2020}.
In contrast, Lagrangian frameworks offer the inherent advantage of determining particles trajectory individually, which can be a key ingredient
for successful design of relevant systems, 
for example, in various microfluidic devices for particle separation.

The Lagrangian framework can be used together with point-wise particles, wherein 
the fluid flow motion equations are solved 
while the effects of particle on fluid flow dynamics are either ignored (one-way coupling) \cite{sato_numerical_2019, olfat_particle_2023} or modeled as source terms in transport equations of continuous phase (two-way coupling) \cite{horwitz_accurate_2016}. 
The particles trajectory is simulated by applying Newton's second law and considering relevant forces acting on particles, \textit{e.g.}, drag force, gravity, etc., which can be fully determined in conjunction with the continuous phase solution.
The Lagrangian approach in conjunction with the assumption of point-wise particles can provide a computationally efficient framework for particle tracking \cite{trunk_development_2021,wang_gpu-accelerated_2022,baldan_efficient_2023}. 
However, they can face challenges to model particle-wall interactions accurately, which can be problematic in applications that these interactions play a key role in dictating the dynamics of particle and/or fluid flow.

The design and optimization of many passive and label-free microfluidic devices require accurate prediction of fluid flow and particle trajectory, necessitating particle-wall interactions to be resolved precisely.
An appropriately-designed geometrical configuration can enable various passive mechanisms for size-dependent particle separation, which is a key component of lab-on-a-chip systems for diagnostic purposes \cite{tang_geometric_2022}.
For example, with regard to the wide spectrum of hydrodynamic filtration methods such as deterministic lateral displacement (DLD) \cite{huang_continuous_2004,liang_scaling_2020}, pinched flow \cite{yamada_pinched_2004,takagi_continuous_2005,yamada_hydrodynamic_2005,yang_microfluidic_2006, timary_shear-enhanced_2023}, etc., 
the core principle is to design and produce desirable hydrodynamic characteristics, \textit{e.g.}, pressure and velocity fields,
which in conjunction with complex particle-wall interactions
lead to a set of distinct size-dependent particle trajectories, 
facilitating the separation of particles that are much smaller than minimum opening size of channels, 
as opposed to physical filtration techniques \cite{ji_silicon-based_2008}, which aim at trapping particles that are equal to or larger than pore size.
In order to design and optimize these microfluidic devices, particle-wall interactions need to be captured accurately.

There are multiple particle tracking algorithms/packages available for 
general purposes \cite{macpherson_particle_2009, greifzu_assessment_2016,weinhart_fast_2020}
as well as for specialized applications such as 
fluidized bed
\cite{fries_demcfd_2011, jajcevic_large-scale_2013, deb_novel_2013, capecelatro_eulerlagrange_2013},
general gas-solid flows \cite{garg_open-source_2012} with emphasis on sediment transport
\cite{sun_sedifoam_2016},
porous media \cite{trunk_development_2021},
etc. 
It is also feasible to explicitly model the geometry of 
rigid \cite{atay_dcelectrokinetic_2022} or deforming \cite{villone_simulations_2014, balcazar_multiple_2015, balcazar_level-set_2016} particle
and 
capture the fluid-particle-wall interactions
while solving the fluid flow equations of motion
at a noticeably higher computational cost.
The discrete element method (DEM), which is the underlying technique in some of the aforementioned works, is a popular approach to model granular materials as clusters of rigid particles that interact with each other and potentially with obstacles via contact forces and torques \cite{rojek_contact_2018}.
For example, it can be used to model soil and rocks as particles packed together to study the microstructure and mechanical properties of these materials \cite{liu_matdem-fast_2021}.
It is also used in pharmaceutical \cite{singh_challenges_2022} and agricultural \cite{zhou_modelling_2014} industries among others.
DEM can also be combined with continuum based methods to model fluid-particle interaction \cite{cheng_concurrent_2023}.

Despite these advantages, accurate DEM simulations typically require careful calibration of particle properties and contact models to match real-world behavior, which can be challenging \cite{ye_calibration_2019,coetzee_calibration_2020}.
The calibration is often an iterative process, aiming at identifying a parameter set leading to a resulting system response that matches the bulk behaviour (macro scale) of the system.
The calibration challenge can be intensified by the \textit{ambiguity} of the parameter sets, \textit{i.e.},
there can be a variety of contact parameter sets mapping the same macroscopic reference value. 
One may optimize multiple objective functions simultaneously to take into account the ambiguity of the parameter sets \cite{roessler_development_2019,richter_development_2020}.

To the best of our knowledge, there are currently a few reports in the microfluidics literature attempting to capture the particle-wall interactions \cite{wang_mopsa_2017, ebadi_novel_2019}, 
which are limited to Stokes flow regime, \textit{i.e.}, St$\ll$1, wherein St is the Stokes number.
As these algorithms are developed for Stokes flow regime, 
particles are forced to follow fluid flow streamlines.
As a result, 
correct particle trajectory may not be obtained for relatively high Stokes number  applications, which can occur 
in presence of
sufficiently high-inertia fluid flow, \textit{i.e.}, high Reynolds regime, 
and/or  
particles with sufficiently higher density than that of working fluid.
In addition, some of these works introduce spurious mechanisms and parameters that need to be tuned for different cases in order to produce desirable results \cite{wang_mopsa_2017}.
As a correct solution is typically unknown prior to solving a practical particle tracking problem, 
the algorithm provided by Wang \textit{et. al.} \cite{wang_mopsa_2017} may not be applicable to real-life problems conveniently.

In this paper, we present a new particle tracking algorithm which benefits from the low computational cost of one-way coupling paradigm, 
while particle geometry and its interactions with wall are captured explicitly. 
Our algorithm can be configured to automatically switch between low- and high-Stokes regimes, wherein accurate integration schemes are applied accordingly in a computationally-efficient fashion.
In addition, our method includes a modified variant of  auxiliary structured grid (ASG) technique, in conjunction with a robust mechanism to model particle-wall interactions, which allows determining hosting cell of particles efficiently for unstructured grids with various levels of refinement and/or quality.
We evaluate the robustness and performance of our method by 
studying several benchmarks including microfluidic systems based on DLD and pinched flow.
We also examine the computational cost of both original and modified variants of ASG methods for locating hosting cells.

The remainder of this paper is organized as follows. 
In Section~\ref{sec_method}, the developed particle tracking algorithm is described. 
In Section~\ref{sec_res}, we present our results and discussion.
We conclude in Section~\ref{sec_conclusion} with a brief summary.

%% file: method.tex
\section{Method}
\label{sec_method}
In this work, we use Ansys Fluent software to obtain the solution of fluid flow equations which will be stored on unstructured grids.
In order to examine the robustness of algorithm for complex grids, we use various grids consisting of triangular or hybrid finite cells, with or without inflation layers to apply local mesh refinement near walls.
After obtaining the continuous phase solution, a set of particles is released into fluid flow from one or more given locations and particles trajectory is obtained according to the developed particle tracking algorithm.

\begin{algorithm}[p]
	\caption{The particle tracking algorithm.
	}\label{alg:pt}
	\DontPrintSemicolon
	\SetNoFillComment 
	\KwIn{\\
		\Indp \Indp
		Fluid flow solution \hfill \textcolor{blue}{Fields of velocity ($\vec{V}_f$), density ($\rho_f$), and viscosity ($\mu_f$) stored on a grid}\\
		Particles properties \hfill \textcolor{blue}{Diameter ($d_p$) and density ($\rho_p$) of each particle}\\
		\textit{\textbf{int}} $N_p$ \hfill
		\textcolor{blue}{Number of particles}\\
		\textit{\textbf{double}} $L_\text{ref}$ \hfill
		\textcolor{blue}{Characteristic length of domain}\\
		\textit{\textbf{double}} $t_{max}$ \hfill \textcolor{blue}{Maximum time duration of tracking each particle}\\
		\textit{\textbf{double}} $\Delta t^\ast$ \hfill \textcolor{blue}{Dimensionless time step to enable integration with adaptive time step}\\
		\textit{\textbf{double}} $\text{St}_{p}^\ast$ \hfill \textcolor{blue}{Stokes numbers smaller than this threshold can be considered small}
	}
	\For{$i\gets1$ \KwTo $N_p$}{
		$t\gets 0$ \tcp*{Reset the elapsed time}
		\While{$t< t_\text{max}$}{
			$\vec{r}_p^\text{\hspace{0.2em}n}\gets$ Current position of particle\;
			$\vec{V}_p^\text{n}\gets$ Current velocity of particle\;
			\If{t=0}
			{
				Locate the hosting cell ($C_h$) for the center of particle ($\vec{r}_p^\text{\hspace{0.2em}n}$)\;
				Evaluate the local fluid flow properties, \textit{i.e.}, velocity, density, and viscosity, by interpolating the corresponding values from nodes of hosting cell ($C_h$)\;
			}
			Calculate the particle relaxation time ($\tau_p$) \tcp*{$\tau_p=\frac{\rho_pd_p^2}{18\mu_f}$}
			Calculate the characteristic fluid flow relaxation time ($\tau_{f}$) based on the characteristic length ($L_\text{ref}$) and local velocity magnitude ($U_\text{local}$) of fluid flow \tcp*{$\tau_{f}=\frac{L_\text{ref}}{U_\text{local}}$}
			Calculate the Stokes number of particle ($\text{St}_{p}$) \tcp*{$\text{St}_{p}=\frac{\tau_p}{\tau_{f}}$}
						
			$\Delta t=\Delta t^\ast\times\tau_f$ \tcp*{dimensionless $\Delta t^\ast$ enables integration with an adaptive time step} 		
			
			\eIf{$\text{St}_{p} < \text{St}_{p}^\ast$}
			{
				\tcc{Using Euler scheme, fourth-order Runge-Kutta scheme, etc.}
				Integrate one time step and find $\vec{r}_p^\text{\hspace{0.2em}n+1}$ 
			}
			{
				\tcc{Velocity Verlet time integration algorithm: Part 1}
				Calculate the forces exerted on particle and find the particle acceleration $\vec{a}_p^\text{\hspace{0.2em}n}$\;
				$\vec{V}_p^\text{\hspace{0.2em}n+1/2}\gets \vec{V}_p^\text{\hspace{0.2em}n}+\frac{1}{2}\vec{a}_p^\text{\hspace{0.2em}n}\Delta t$\;
				$\vec{r}_p^\text{\hspace{0.2em}n+1}\gets \vec{r}_p^\text{\hspace{0.2em}n}+\vec{V}_p^\text{\hspace{0.2em}n+1/2}\Delta t$\;
			}
			
			Apply wall-particle interactions as needed\;
			Locate the hosting cell $C_h$ for the new position of particle ($\vec{r}_p^\text{\hspace{0.2em}n+1}$)\;
			Evaluate the local fluid flow properties for the new position ($\vec{r}_p^\text{\hspace{0.2em}n+1}$)\;
			\eIf{$\text{St}_{p} < \text{St}_{p}^\ast$}
			{
				$\vec{V}_p^\text{n+1}\gets \vec{V}_f$ \tcp*{set the particle velocity as that of local fluid flow at $\vec{r}_p^{\hspace{0.2em}n+1}$} 
			}
			{
				\tcc{Velocity Verlet time integration algorithm: Part 2}
				Calculate forces exerted on the particle and find the particle's new acceleration $\vec{a}_p^\text{\hspace{0.2em}n+1}$\;
				$\vec{V}_p^\text{\hspace{0.2em}n+1}\gets \vec{V}_p^\text{\hspace{0.2em}n+1/2}+\frac{1}{2}\vec{a}_p^\text{\hspace{0.2em}n+1}\Delta t$\;
			}
			$t\gets t+\Delta t$\;
		}
	}
\end{algorithm}

Algorithm~\ref{alg:pt} provides a pseudo-code for the particle tracking method used in this work.
Our algorithm can simulate trajectory of point-wise particles as well as those with a nonzero finite size.
In this work, our focus is on the latter case which is particularly important when particle size plays a key role in particle-wall interactions, \textit{e.g.}, hydrodynamic separation techniques including DLD, pinched flow, etc.

The computational cost of Algorithm~\ref{alg:pt} is dominated by the task of locating hosting cells, \textit{i.e.}, a cell encompassing a given position, which is applied prior to interpolation of fluid flow properties (line 22) and more heavily when capturing wall-particle interactions (line 21).
The reason for more expensive computational cost of the latter is that 
the number of times that a hosting cell needs to be located increases with the number of vertices considered on perimeter of particle as discussed in more detail in Section~\ref{sec_particle_wall_interactions}.
The key components of Algorithm~\ref{alg:pt} are described in more detail in the following.

\subsection{Locating hosting cell}
There are multiple different techniques for determining hosting cell.
A simple approach is to use an auxiliary structured grid (ASG) to determine unstructured cells (from the main unstructured mesh) intersecting each cell of ASG. 
The process is performed only once and prior to the main particle tracking simulation. The information stored on ASG is used during the cell locating process when running particle tracking simulation. 
The ASG provides a relatively short list of unstructured cells to search over for locating hosting cell instead of searching over the entire cells of unstructured grid.

A disadvantage of ASG approach is that it can become inefficient when the unstructured mesh is highly non-uniform. For example, some cells of ASG may overlap a very large number of unstructured cells in areas that the unstructured mesh is refined, \textit{e.g.}, near walls where fluid-wall interactions need to be captured accurately. 
There are mainly two classes of techniques aiming at resolving the aforementioned issue: hierarchical tree (HT) \cite{lubbe_analysis_2020}, and various neighbor search (NS) algorithms \cite{lohner_vectorized_1990, haselbacher_efficient_2007, macpherson_particle_2009, zhao_leveraging_2023}.
The HT method is an efficient technique to locate hosting cell, but it introduces additional memory and computational costs compared to ASG-based methods. It is also relatively complex to implement \cite{wang_gpu-accelerated_2022}.

The NS method is an efficient technique to locate hosting cell.
The core idea behind NS is to inspect last segment of particle trajectory (LSPT), which connects previous position of particle to its current location. 
Assuming that hosting cell for previous position is known, the configuration of LSPT together with the most recent hosting cell can be examined to determine which side of the most recent hosting cell may have been intersected by LSPT.
In case none of the sides of most recent hosting cell is crossed, the hosting cell remains the same and no hosting cell update is needed.
In the case that a side of the most recent hosting cell is intersected, the corresponding neighboring cell is determined and is considered as the updated hosting cell. 
The process is repeated for the updated hosting cell until no side of the most recent hosting cell is intersected by LSPT.

A disadvantage of NS is that it can become inefficient in the case that LSPT is noticeably longer than the size of local unstructured cells, which can happen where unstructured mesh is very fine and/or integration time-step for particle tracking is relatively large.
Under these circumstances, a particle can cross several cells in one time-step resulting in a relatively large computational cost to locate hosting cell.
In addition, NS is not self-starting. 
That is, as there is no previous hosting cell at the beginning of simulation (first time-step), one has to resort to other methods to find the initial hosting cell to start the particle tracking.

In this work, in addition to our developed tracking algorithm for finite-size particles, 
we report a new two-step search algorithm to mitigate the inefficiency associated with the original ASG method.
In our search algorithm, each structured cell of ASG stores information related to nodes/cells of unstructured grid that reside inside the said structured cell.
To find the hosting cell of a particle, 
first, the closest node to the particle is found by searching over the shortlisted nodes in proximity of the particle.
In the second step, an outward spiral search is conducted over the cells around the said closest node to find the hosting cell of particle.

\begin{figure}[!bt]
\centering
\subfloat[]{\includegraphics[width=0.45\textwidth, cfbox=blue 2pt 0pt]{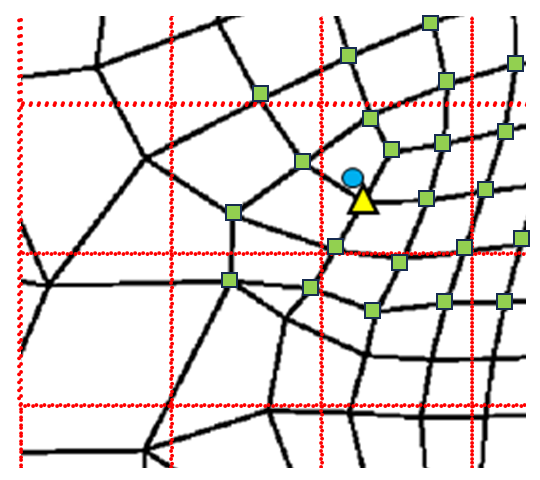}}
\qquad
\subfloat[]{\includegraphics[width=0.45\textwidth, cfbox=blue 2pt 0pt]{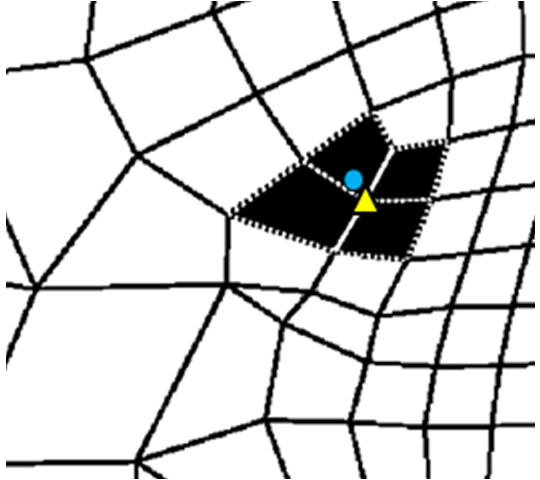}}
\caption{
A schematic representation of search algorithm to locate hosting cell, including 
an exemplary hybrid unstructured grid (solid lines) to store fluid flow properties, \textit{e.g.} pressure, velocity, density, viscosity, etc., together with an auxiliary structured grid (dotted lines) used to store information related to nodes/cells of unstructured grid 
residing inside structured cells.
The process of locating hosting cell for a given particle, shown as a circle here, consists of
1. searching over all shortlisted nodes near particle (shown as square/triangle), and finding closest node (shown as triangle) to the particle (a), 
and
2. searching over a stack of cells around the closest node (filled cells) 
to find hosting cell (b).
There are examples of multi-level search in an outward spiral manner depicted in Fig.~\ref{fig_cell_host_case_01}.
}
\label{fig_cell_host_case_00}
\end{figure}

This process is illustrated in Fig.~\ref{fig_cell_host_case_00} with an example.
As shown in Fig.~\ref{fig_cell_host_case_00} (a), the closest node (shown as a triangle) to a given particle (shown as a circle) is first determined.
A search is, then, performed over the first layer of cells (filled cells) around the closest node to locate hosting cell; Fig.~\ref{fig_cell_host_case_00} (b).
The advantage of this approach compared to a simple ASG-based method is that instead of searching over the entire list of cells within a structured cell of ASG, and checking whether any of those cells encompasses the particle location, a search is performed only on a small number of cells in proximity of the closest node to the particle.
It is worth mentioning that in this modified technique, determining the ``closest node'' itself requires a search over the entire list of nodes stored by the corresponding structured cell(s) of ASG.
However, the key point is that depending on configurations, there are situations wherein the vectorized calculation of distance between a set of points can become less expensive than determining whether a point falls inside a set of polygons.
For these situations, the modified version of ASG can reduce the cost of locating hosting cell.

In addition to being self starting,
our modified ASG technique offers the inherent ASG advantage of locally searching for hosting cell regardless of previous position of particle.
As a result, very fine unstructured grids together with relatively large integration time steps can be utilized as needed without incurring additional costs that would otherwise need to be dealt with when using NS approaches 
due to particle crossing a large number of unstructured cells.

\begin{figure}[!bt]
\centering
\subfloat[]{\includegraphics[width=0.31\textwidth, cfbox=blue 2pt 0pt]{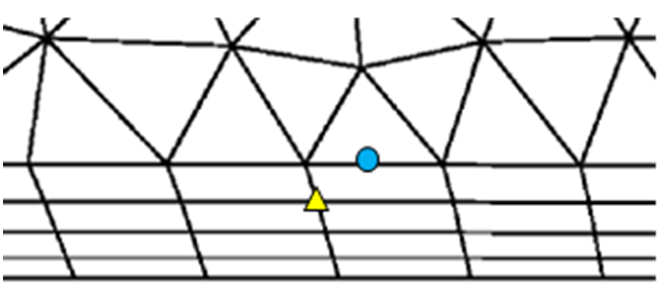}}
\hspace{0.55em}
\subfloat[]{\includegraphics[width=0.31\textwidth, cfbox=blue 2pt 0pt]{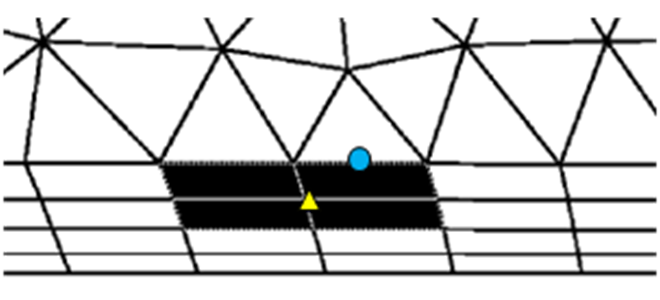}}
\hspace{0.55em}
\subfloat[]{\includegraphics[width=0.31\textwidth, cfbox=blue 2pt 0pt]{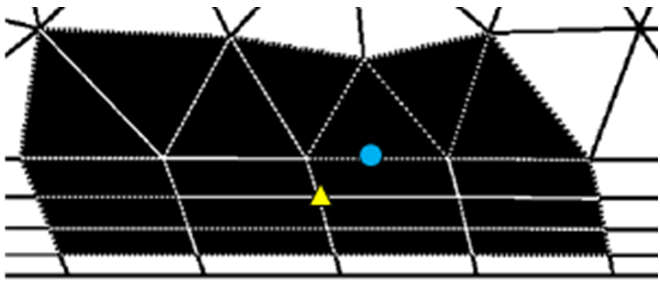}}
\hfill
\vspace{1em}
\subfloat[]{\includegraphics[width=0.23\textwidth, cfbox=blue 2pt 0pt]{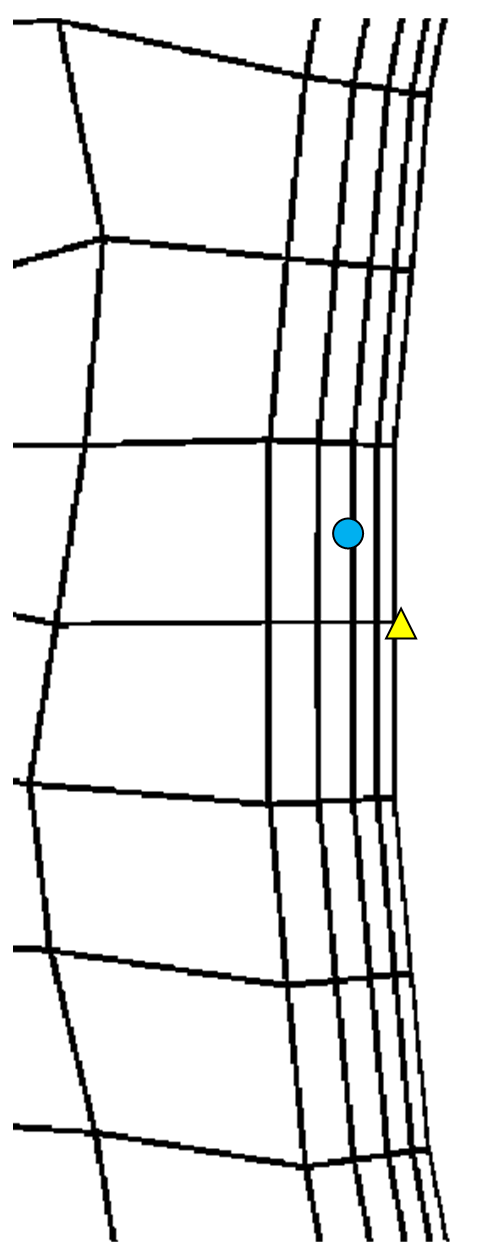}}
\hspace{0.25em}
\subfloat[]{\includegraphics[width=0.23\textwidth, cfbox=blue 2pt 0pt]{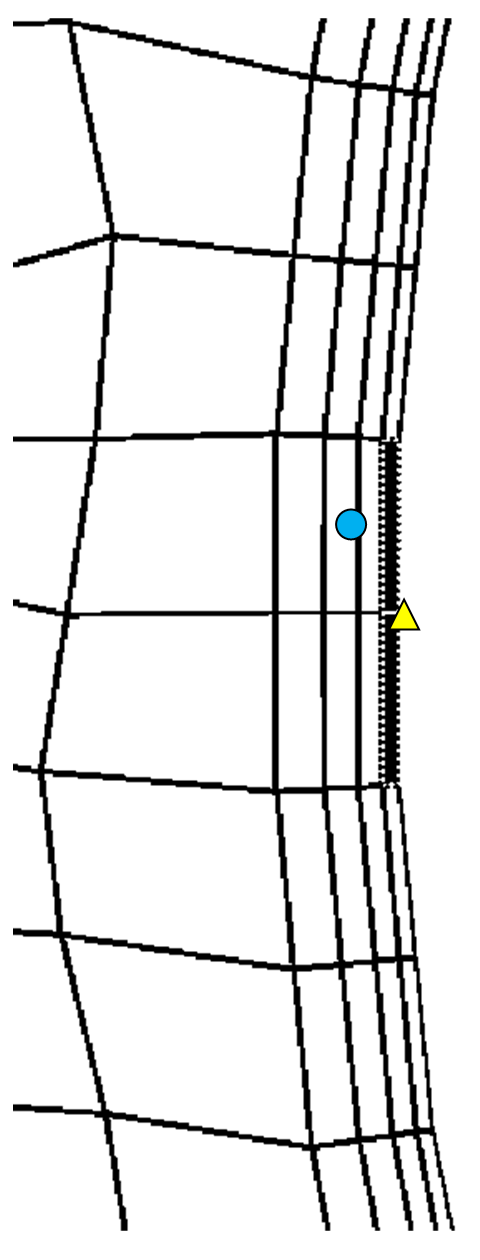}}
\hspace{0.25em}
\subfloat[]{\includegraphics[width=0.23\textwidth, cfbox=blue 2pt 0pt]{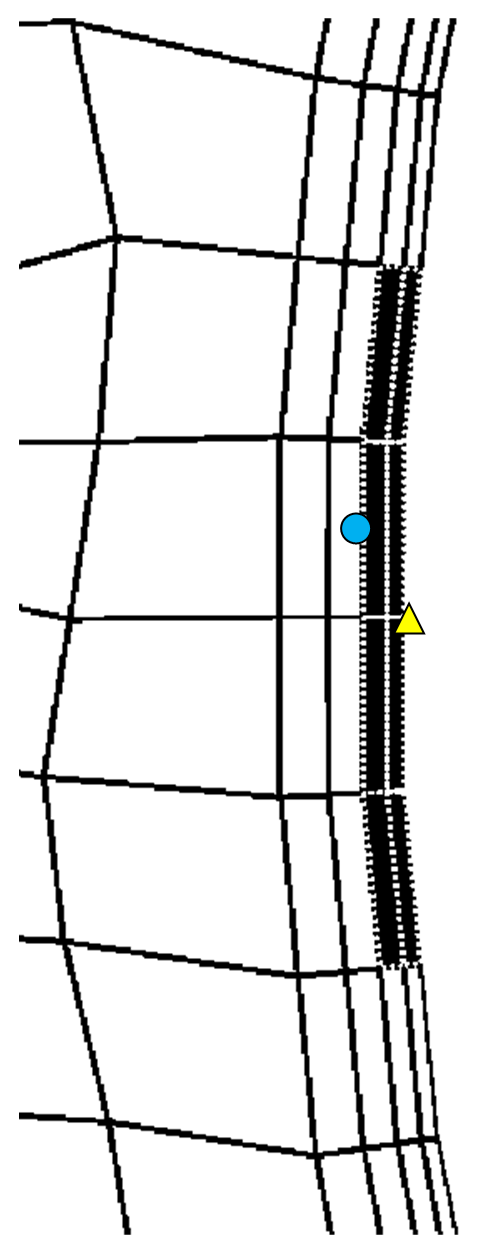}}
\hspace{0.25em}
\subfloat[]{\includegraphics[width=0.23\textwidth, cfbox=blue 2pt 0pt]{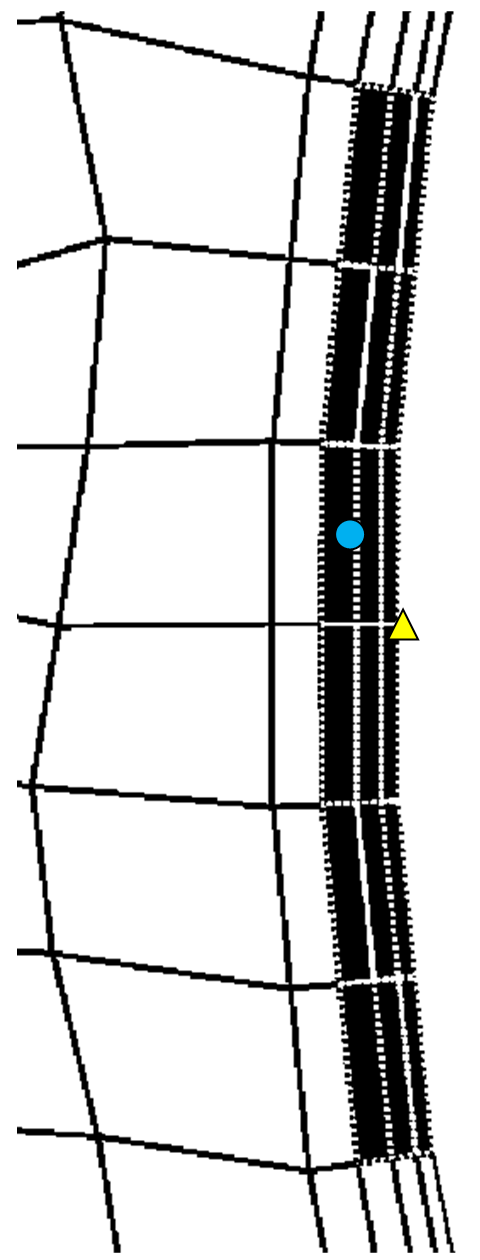}}
\caption{
A schematic representation of our search algorithm with an outward spiral scheme to locate hosting cell for two exemplary cases (a\textendash c and d\textendash g).
After determining closest node (shown as triangle) to a given particle (shown as circle) (a) and (d),
a search is performed over the first stack of cells (filled cells) around the closest node (b) and (e).
If hosting cell is not found, the second stack of cells is considered for searching hosting cell (c) and (f).
If hosting cell is not found, the third stack of cells will be considered for searching hosting cell (g).
Considering a maximum of 3\textendash 4 stacks of cells proved to be sufficient for very low-quality grids intentionally made to consist of highly-skewed cells, and/or inflation layers with high aspect ratio cells.
Not locating a hosting cell after searching through a reasonably large number of stacks of cells is an indication of particle being outside of computational domain.
}
\label{fig_cell_host_case_01}
\end{figure}

In a high-quality mesh, the second step of the search algorithm includes only the core cells, directly connected to the closest node, as the candidate cells, \textit{i.e.}, no additional stack of cells needs to be considered ($\Phi=0$).
In a practical mesh, where the unstructured cells can be highly skewed, and/or with extremely high/low aspect ratios, there are scenarios that hosting cell may not be directly connected to closest node of particle.
A couple of exemplary cases are illustrated in Fig.~\ref{fig_cell_host_case_01}.
In order to deal with such low-quality cells, which can be present in a practical grid, we consider a reasonably large number of additional stacks of cells $\Phi$, \textit{e.g.}, 1, 2, 3 or 4, to be searched over in an outward spiral manner.
Not finding a hosting cell is an indication that the particle has crossed a wall boundary.
It should be noted that the additional stacks are examined only as needed, \textit{i.e.}, when the hosting cell is not available in the already-examined stacks.
In addition, once the hosting cell is found, the search is terminated.
These measures aim at minimizing the additional costs related to adding stacks of cells when dealing with low-quality unstructured grids.

\subsection{Particle-wall interactions}
\label{sec_particle_wall_interactions}
\begin{figure}[!bt]
\centering
\includegraphics[width=0.5\textwidth, cfbox=blue 2pt 0pt]{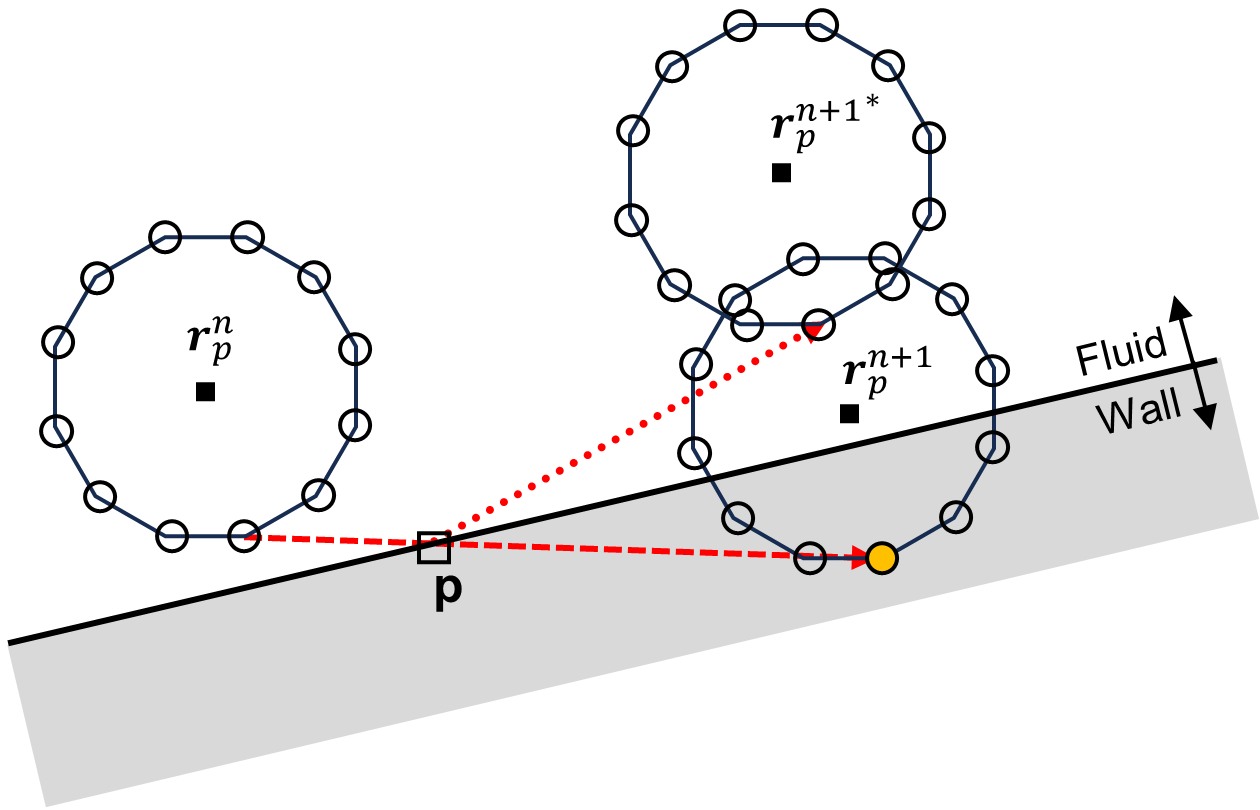}
\caption{
A schematic representation of particle-wall interactions, wherein a specular reflection scheme is used to reflect a particle with a nonzero finite size with previous location of $\boldsymbol{r}_p^{n}$ and current position of $\boldsymbol{r}_p^{n+1}$ off a wall. 
The perimeter of particle is represented by a finite number of circumferential points (12 in this example).
The deepest penetration of circumferential points into the wall is determined (the corresponding circumferential point is shown as filled circle here) and used for determining the position of circumferential point after reflection.
The same net translation transformation is applied to the entire particle afterwards to update its position ($\boldsymbol{r}_p^{n+1^\ast}$).
}
\label{fig_bc}
\end{figure}

In this work, a particle is represented by a circular solid object with a finite number of circumferential points.
A specular reflection scheme is used to reflect a particle off a wall boundary.
A schematic representation of this mechanism is shown in Fig.~\ref{fig_bc}.
In order to reflect a particle that has crossed a wall boundary partially or entirely back into fluid domain, a search is performed over the circumferential points of particle and the depth of penetration into wall is calculated for each circumferential point.
The circumferential point with the deepest penetration (shown as a filled circle in Fig.~\ref{fig_bc}) is then used to find the translation vector $\boldsymbol{r}_p^{n+1^\ast}-\boldsymbol{r}_p^{n+1}$ to be applied to the particle and to determine the updated position of particle ($\boldsymbol{r}_p^{n+1^\ast}$).

In addition to updating the position of particle when bouncing off a wall boundary, we also update the particle velocity according to Eq.~\ref{eq-bc-vel},  
\begin{equation}
\vec{V}_p^\text{\hspace{0.2em}post}=\vec{V}_p^\text{\hspace{0.2em}prior}-2\alpha V_n\hat{n},
\label{eq-bc-vel}
\end{equation}
wherein $\vec{V}_p^\text{\hspace{0.2em}prior}$ and $\vec{V}_p^\text{\hspace{0.2em}post}$ refer to the particle velocity prior to and post reflection, respectively,
$V_n$ denotes the component of particle velocity normal to the wall,
$\hat{n}$ shows the unit normal vector of wall boundary (pointing towards the fluid domain),
and $\alpha$ is the elastic reflection coefficient, 
wherein $\alpha=0$ corresponds to a pure inelastic reflection, 
$\alpha=1$ corresponds to a pure elastic reflection, 
and $0<\alpha<1$ is related to a partially-elastic reflection.

\subsection{Time integration}
An adaptive time-step integration is implemented in Algorithm~\ref{alg:pt}, wherein an appropriate time step is calculated based on the Stokes number of particle.
For a sufficiently small Stokes number, we assume that the particle instantaneously follows the local streamline of fluid flow, while for relatively large Stokes numbers, a velocity Verlet algorithm is implemented to update the instantaneous acceleration, velocity, and position of particle.

\subsubsection{Small Stokes regime}
The Stokes number is defined as the ratio of particle relaxation time, $\tau_p={\rho_pd_p^2}/{18\mu_f}$, 
to that of fluid flow, 
$\tau_f=L_\text{ref}/U_\text{local}$, \textit{i.e.}, $\text{St}_p\equiv{\tau_p}/{\tau_f}={\rho_pd_p^2U_\text{local}}/{18\mu_fL_\text{ref}}$, wherein $\rho_p$ refers to the particle density, $d_p$ denotes the particle diameter, $\mu_f$ is the fluid viscosity, $L_\text{ref}$ shows the characteristic length of system (minimum feature size), and $U_\text{local}$ is the magnitude of local fluid flow velocity, respectively.
For the case that Stokes number is sufficiently small ($\text{St}_p\ll 1$), the particle can be assumed to follow the local streamline instantaneously.
This allows the use of a much larger integration time-step (of order of fluid flow characteristic relaxation time) than that one would have to use by integrating the acceleration-dependent equations of motion (of order of particle relaxation time).

It is also worth mentioning that using the local velocity magnitude 
provides a more accurate measure of 
how large the particle relaxation time is compared to that of fluid flow locally (local Stokes number),
allowing larger time steps to be applied safely, 
as opposed to using a fixed characteristic velocity magnitude for the system, \textit{e.g.}, maximum velocity magnitude throughout the domain, which would lead to applying an excessively small time step for some parts of simulation inefficiently.

For the case of small Stokes number, we can use one of numerous integration schemes to update the position of particle without calculating the net forces acting on particle.
Then, the particle velocity can be updated to match that of fluid flow at new location.
Here, we implement only two methods: explicit Euler scheme, and fourth-order Runge-Kutta technique.
The Euler scheme can be described as:
\begin{equation}
\vec{r}_p^{\hspace{0.2em}n+1}=\vec{r}_p^{\hspace{0.2em}n}+\Delta t\vec{V}_p^{n},
\label{eq-euler}
\end{equation}
wherein $\vec{r}_p^{\hspace{0.2em}n}$ and $\vec{r}_p^{\hspace{0.2em}n+1}$ denote the position of particle at two consecutive time steps $t$ and $t+\Delta t$, respectively, while $\vec{V}_p^{n}$ is the velocity of particle at time $t$.
The truncation error of Euler method is of order of $\Delta t^2$.

The fourth-order Runge-Kutta method  
with a truncation error of order of $\Delta t^5$
may be described as shown in Eq.~\ref{eq-rk4}:
\begin{equation}
\vec{r}_p^{\hspace{0.2em}n+1}=\vec{r}_p^{\hspace{0.2em}n}+\frac{\Delta t}{6}\big(
\vec{V}_p^{n} + 
2\vec{V}_p^{n^\ast} + 
2\vec{V}_p^{n^{\ast\ast}} + 
\vec{V}_p^{n^{\ast\ast\ast}}
\big),
\label{eq-rk4}
\end{equation}
wherein 
$\vec{V}_p^{n^\ast}$, $\vec{V}_p^{n^{\ast\ast}}$, and $\vec{V}_p^{n^{\ast\ast\ast}}$ denote the fluid flow velocity at positions $\vec{r}_p^{\hspace{0.2em}n^\ast}$, $\vec{r}_p^{\hspace{0.2em}n^{\ast\ast}}$, and $\vec{r}_p^{\hspace{0.2em}n^{\ast\ast\ast}}$, respectively,
in which
\begin{equation}
\begin{split}
\vec{r}_p^{\hspace{0.2em}n^\ast}&=\vec{r}_p^{\hspace{0.2em}n}+\frac{\Delta t}{2}\vec{V}_p^{\hspace{0.2em}n}\\
\vec{r}_p^{\hspace{0.2em}n^{\ast\ast}}&=\vec{r}_p^{\hspace{0.2em}n}+\frac{\Delta t}{2}\vec{V}_p^{\hspace{0.2em}n^\ast}\\
\vec{r}_p^{\hspace{0.2em}n^{\ast\ast\ast}}&=\vec{r}_p^{\hspace{0.2em}n}+\Delta t\vec{V}_p^{\hspace{0.2em}n^{\ast\ast}}\\
\end{split}
\label{eq-rk4-mid_pos}
\end{equation}

\subsubsection{Large Stokes regime}
The acceleration of particle, $\vec{a}_p$, is obtained from the total force per unit mass exerted on particle as:
\begin{equation}
\vec{a}_p=\sum\vec{f}=\vec{f}_D+\vec{f}_g+\vec{f}_\text{ext}+\cdots,
\label{eq-acc}
\end{equation}
wherein
$\vec{f}_D$ is drag force per unit mass of particle exerted by working fluid,
$\vec{f}_g$ is gravitational acceleration, 
$\vec{f}_\text{ext}$ is any additional external force per unit mass due to electric field, magnetic field, etc.
There are also other forces that can be considered based on application needs, \textit{e.g.}, 
Brownian force,
thermophoretic force,
Saffman force,
Magnus force,
etc. \cite{mahian_recent_2019}.
The drag force per unit mass of particle, $\vec{f}_D$, can be obtained from Eq.~\ref{eq-drag} as
\begin{equation}
\vec{f}_D=\frac{3\mu_fC_D\text{Re}_p}{4\rho_pd_p^2}\big(\vec{V}_f-\vec{V}_p\big),
\label{eq-drag}
\end{equation}
wherein 
$\vec{V}_f$ denotes local velocity of fluid flow,
$\vec{V}_p$ is particle velocity,
$\text{Re}_p\equiv\rho_f\|\vec{V}_s\|d_p/\mu_f$ is particle Reynolds number,
in which $\vec{V}_s=\vec{V}_f-\vec{V}_p$ is slip velocity of particle,
and $\rho_f$ denotes fluid density.
In addition, the drag coefficient, $C_D$, is given by Eq.~\ref{eq-cd} as 
\begin{equation}
C_D=\frac{24}{\text{Re}_p}\big[1+0.15\text{Re}_p^{0.687}\big],
\label{eq-cd}
\end{equation}
which is valid for a relatively wide range of Reynolds number, 
spanning from Stokes flow ($\text{Re}_p\ll1$) to transition regime with an upper limit of $\text{Re}_p\sim800$ \cite{ahmadi_transport_2008}.

%% file: results.tex
\section{Results}
\label{sec_res}
In this section, we report the experimental and numerical results of particle tracking studies for a microfluidic DLD device in Section~\ref{res_dld}.
We also present our findings from modeling particle separation based on the pinched flow method and compare our results with those from an experimental work reported by Yamada \textit{et. al.} \cite{yamada_pinched_2004} in Section~\ref{res_pinched_flow}.
We also characterize the computational performance of our algorithm and its implementation in C++ language in Section~\ref{sec_res_performance_eval}.
In these studies, we use Ansys Fluent package to solve the fluid flow equations of motion. 
We, then, use the obtained velocity field as an input to our developed program for particle tracking simulation based on Algorithm~\ref{alg:pt}.

\input{results_dld}

\input{results_pinched_flow}

\input{results_performance_evaluation}

%% file: results_dld.tex
\subsection{DLD}
\label{res_dld}
\begin{figure*}[!tb]
	\centering
	\includegraphics[width=\textwidth, cfbox=blue 2pt 0pt]{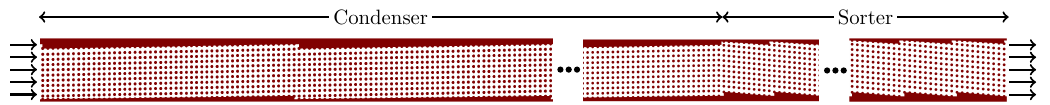}
	\captionof{figure}{
		A schematic representation of microfluidic DLD device with a condenser and sorter design consisting of circular pillar array with diameter and spacing of 14 $\mu m$.
		The inclination of pillar array is 1.15$^\circ$ (periodicity of $N_p=50$) and -5.71$^\circ$ (periodicity of $N_p=10$)
		for condenser and sorter, respectively.
	}
	\label{fig_res_dld_geom}
\end{figure*}

In this section, we study motion of particles inside a microfluidic DLD device with a condenser and sorter array design \cite{kim_broken_2017} experimentally and numerically.
We used our recently developed DLD design automation (DDA) tool \cite{mehboudi_universal_2024,mehboudi_mnflow_2024} to design the DLD layouts.
A schematic representation of our device structure is illustrated in Fig.~\ref{fig_res_dld_geom}.
The device consists of two sets of DLD arrays attached together along the channel axis.
The first array forms a condenser with a small critical diameter $d_{c,1}$, wherein particles larger than $d_{c,1}$ are displaced laterally while moving through the device and are accumulated close to one of the channel sidewalls (accumulation sidewall).
The second array forms a sorter with a relatively large critical diameter $d_{c,2}$, wherein particles larger than $d_{c,2}$ are displaced laterally and are accumulated close to 
the channel sidewall opposite of the condenser accumulation sidewall.

The device consists of circular pillars with diameter and spacing of 14 $\mu m$, while the array of pillars is tilted about 1.15$^\circ$ with respect to the channel axis in the condenser (periodicity of $N_p=50$)
and 
-5.71$^\circ$ in the sorter (periodicity of $N_p=10$).
The critical diameters are predicted to be about $d_{c,1}=3.0~\mu m$ and $d_{c,2}=6.5~\mu m$ according to empirical formula $d_c=1.4g(1/N_p)^{0.48}$ \cite{davis_microfluidic_2008}, in which $g$ denotes gap between adjacent pillars.

\subsubsection{Experimental}
\label{sec_res_dld_exp}
The device was fabricated by using photolithographic etching.
Briefly, the fabrication process flow starts with growing a 1-micron thick silicon dioxide on a silicon wafer previously cleaned in piranha bath.
The channel patterns are defined on silicon wafer through photolithography by using the positive AZ 5209-E photoresist.
The patterns are transferred into silicon dioxide layer by using reactive ion etching (RIE), first, and into silicon wafer by using deep reactive ion etching (DRIE) afterwards.
Next, the photoresist is removed in a piranha bath. 
The silicon dioxide layer is stripped using diluted hydrofluoric acid.
Finally, the silicon wafer is anodically bonded to a glass wafer previously processed by drilling 1-mm holes to access inlet and outlet of device. 

We used polydimethylsiloxane (PDMS) to create fluidic interconnects and bonded them to the glass wafer after a gentle oxygen plasma treatment to facilitate injection and collection of sample into and from the microfluidic device. 
We acquired $1.3~\mu m$ and $5~\mu m$ fluorescent particles from Spherotech Inc.
We used syringe pump to inject sample into device at a flow rate of 1\textendash5 $\mu l$/\text{min} and visualized the particles motion by using a Nikon SMZ25 fluorescence stereoscope.

\subsubsection{Results}
\label{res_dld_discussion}
\begin{figure}[!hbt]
	\centering
	\subfloat[Upstream]{\includegraphics[width=\textwidth, cfbox=blue 2pt 0pt]{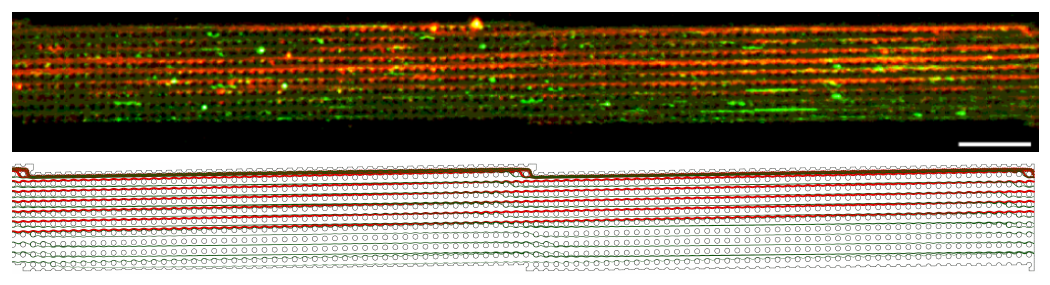}}
	\vspace{1em}
	\subfloat[Downstream]{\includegraphics[width=\textwidth, cfbox=blue 2pt 0pt]{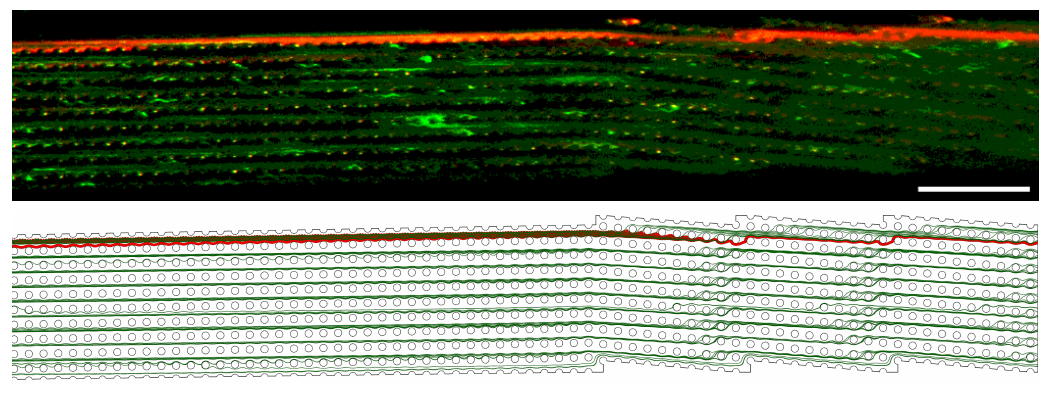}}
	\caption{
		Fluorescence microscopy image together with modeling results of tracking particles 
		of diameter 
		$d=1.3~\mu m$ (green) and $d=5.0~\mu m$ (red)
		within upstream (a) and downstream (b) of microfluidic DLD device.
		Modeling results are obtained by using Algorithm \ref{alg:pt}.
		Large particles show a bumping mode displacement within the condenser ($d>d_{c,1}$), while they have a zig-zag movement within the sorter ($d<d_{c,2}$).
		Small particles have a zig-zag displacement mode throughout the device as they are smaller than both critical diameters $d_{c,1}$ and $d_{c,2}$.
	}
	\label{fig_res_dld_exp_sim}
\end{figure}

The experimental and numerical results obtained for tracking 1.3-$\mu m$ and 5.0-$\mu m$ particles are shown in Fig.~\ref{fig_res_dld_exp_sim}.
The 
It can be discerned that 
small particles have a zig-zag movement mode with a net displacement along the channel axis throughout the device as they are smaller than critical diameters of condenser, $d_{c,1}$, and sorter, $d_{c,2}$.
In contrast, large particles are laterally displaced (bumping mode displacement) 
while moving along the channel within the condenser as they are larger than the corresponding critical diameter $d_{c,1}$.
However, these particle show a zig-zag motion mode within the sorter as they are smaller than the corresponding critical diameter $d_{c,2}$.
The experimental and modeling results match acceptably.

\begin{figure*}[!tb]
	\centering
	\includegraphics[width=\textwidth,cfbox=blue 2pt 0pt]{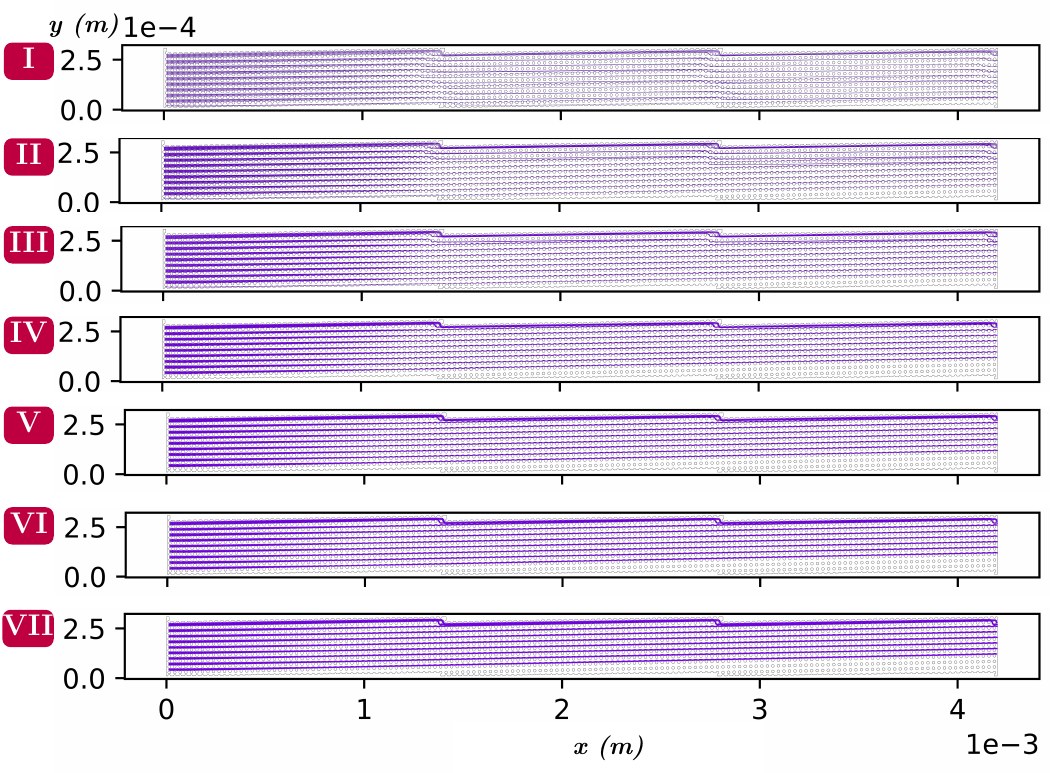}
	\captionof{figure}{
		Modeling results of tracking particles of diameter 
		$1.3~\mu m$ (I),
		$2.0~\mu m$ (II),
		$3.0~\mu m$ (III),
		$4.0~\mu m$ (IV),
		$5.0~\mu m$ (V),
		$6.0~\mu m$ (VI),
		and
		$7.0~\mu m$ (VII) in upstream of device obtained by using Algorithm \ref{alg:pt}.
		A transition from ziz-zag to bumping displacement mode occurs for particles of diameter of $\sim 3.0~\mu m$.
	}
	\label{fig_res_dld_streaklines_us}
\end{figure*}

\begin{figure*}[!tb]
	\centering
	\includegraphics[width=\textwidth,cfbox=blue 2pt 0pt]{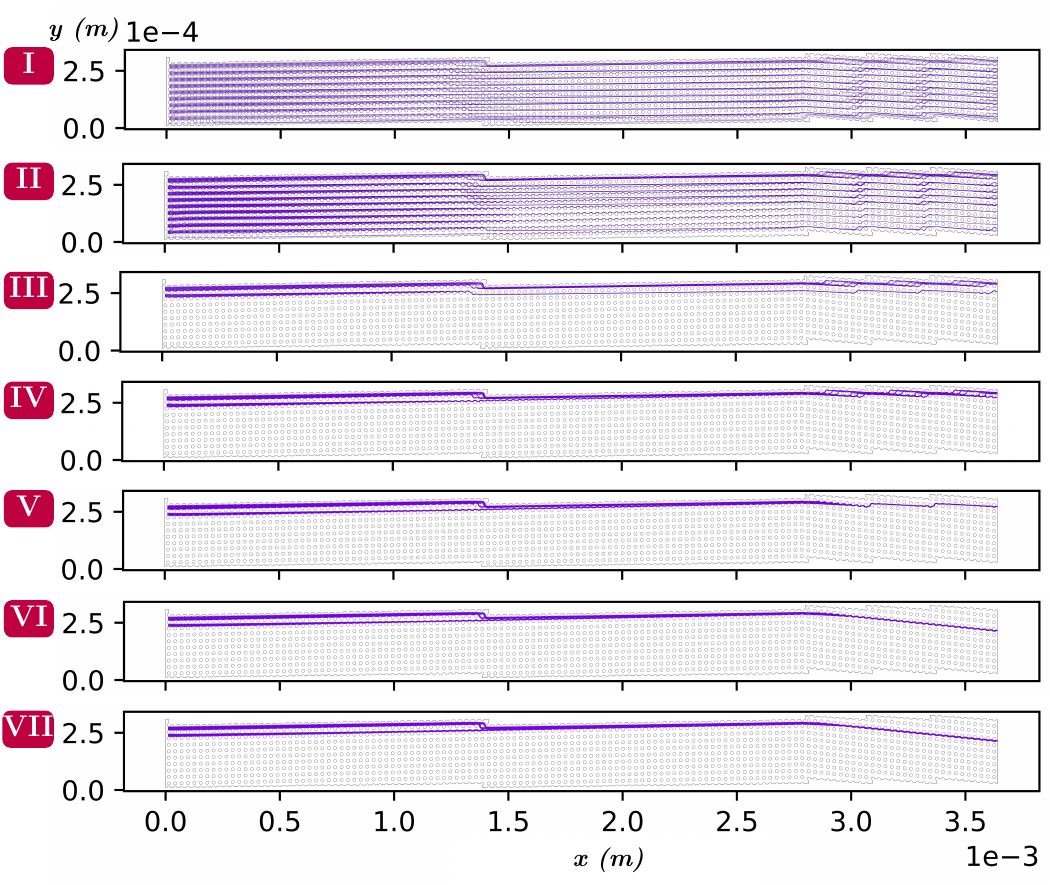}
	\captionof{figure}{
		Modeling results of tracking particles of diameter 
		$1.3~\mu m$ (I),
		$2.0~\mu m$ (II),
		$3.0~\mu m$ (III),
		$4.0~\mu m$ (IV),
		$5.0~\mu m$ (V),
		$6.0~\mu m$ (VI),
		and
		$7.0~\mu m$ (VII) in downstream of device obtained by using Algorithm \ref{alg:pt}.
		A bumping mode in sorter is predicted for particles of $\sim 6~\mu m$ in diameter and larger.
	}
	\label{fig_res_dld_streaklines_ds}
\end{figure*}

We have also modeled motion of particles of various sizes. 
Some of the obtained results are shown in Figs.~\ref{fig_res_dld_streaklines_us} and ~\ref{fig_res_dld_streaklines_ds} for upstream and downstream of device, respectively.
It is discernible from Figs.~\ref{fig_res_dld_streaklines_us} that a transition from zig-zag to bumping mode displacement occurs within the condenser for particles of diameter $\sim 3~\mu m$ which is in accordance with the empirical data reported in the literature \cite{davis_microfluidic_2008}.
In addition, 
Figs.~\ref{fig_res_dld_streaklines_ds} suggests that a transition from zig-zag to bumping mode displacement should take place within the sorter for particles of diameter $\sim 6~\mu m$ which is in acceptable agreement with the value of $\sim 6.5~\mu m$ obtained from the empirical formula reported by Davis \cite{davis_microfluidic_2008}.

%% file: results_pinched_flow.tex

\subsection{Pinched flow}
\label{res_pinched_flow}
In this section, we study a microfluidic chip with pinched flow.
We select one of the devices reported by Yamada \textit{et. al.} \cite{yamada_pinched_2004} as a benchmark to compare the outcome of our modeling with their reported experimental results.
A schematic representation of the microfluidic device is shown in Fig.~\ref{res_pinched_flow}.
\begin{figure*}[!tb]
	\centering
	\includegraphics[width=\textwidth,cfbox=blue 2pt 0pt]{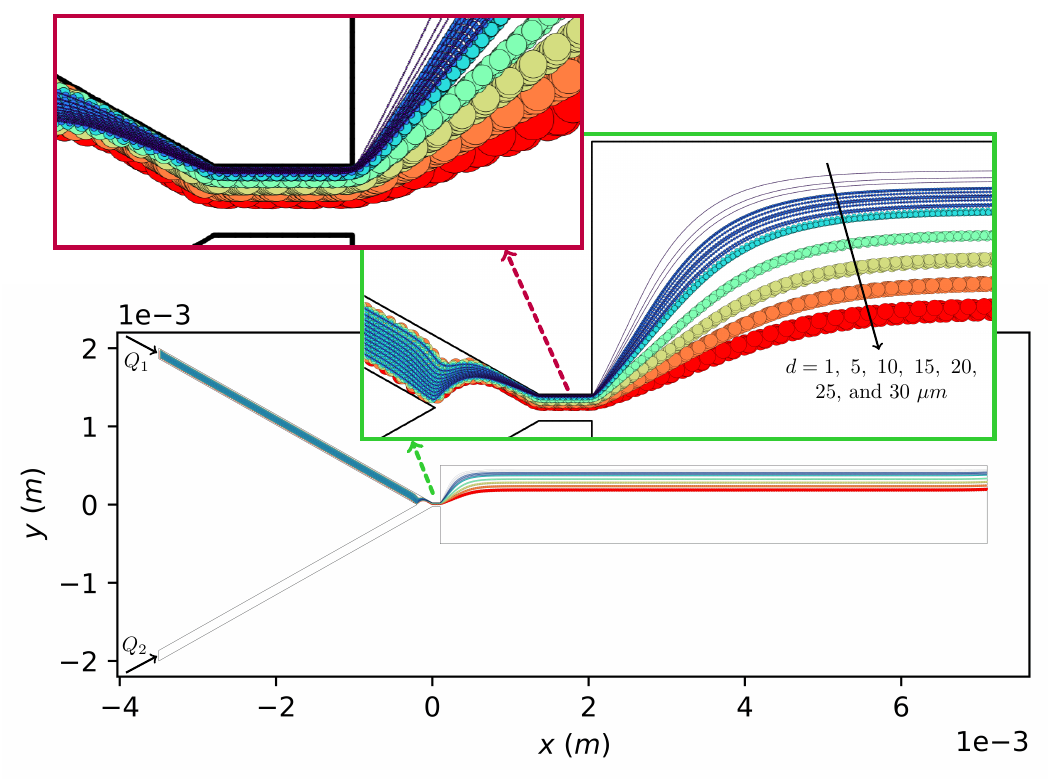}
	\captionof{figure}{
		Modeling results of tracking particles of diameter 
		$1~\mu m$,
		$5~\mu m$,
		$10~\mu m$,
		$15~\mu m$,
		$20~\mu m$,
		$25~\mu m$,
		and
		$30~\mu m$ in a microfluidic device with pinched flow similar to that reported by Yamada \textit{et. al.} \cite{yamada_pinched_2004} obtained by using Algorithm \ref{alg:pt}.
		Width of particle-containing solution in pinched segment is $\sim6~\mu m$.
		Particles with a radius sufficiently close to this value or larger are pressed against and sorted along one of sidewalls in pinched segment enabling them to follow distinct streamlines leading to their separation in expanded segment.
	}
	\label{fig_res_pinched}
\end{figure*}
The pinched segment is $50~\mu m$ wide and $100~\mu m$ long, while the expanded region has a width of $1~mm$ and a length of $7~mm$.
The expansion ratio is defined as ratio of width of expanded region to that in pinched region and equals $\psi=20$ for this structure.
While the volumetric flow rate of particle-containing solution, $Q_1$, and that of sheath flow, $Q_2$, vary to study their effects on separation efficiency, the total volumetric flow rate remains constant and equals 140 $\mu l/h$.

It is worth mentioning that our modeling in this work is two-dimensional (2D), because of which some discrepancies with the experimental work of Yamada \textit{et. al.} \cite{yamada_pinched_2004} are expected.
In particular, in the work of Yamada \textit{et. al.} \cite{yamada_pinched_2004}, the expanded region is relatively shallow ($50~\mu m$ deep compared to its $1~mm$ width), while the pinched segment has a depth-to-width ratio of about unity ($50~\mu m:50~\mu m$).
Herein, we consider a depth of $50~\mu m$ when calculating flow rate. 
However, when solving the fluid flow equations of motions in a 2D fashion, the depth of channels is assumed to be infinitely long.
As a result, the velocity field obtained numerically is not expected to match accurately the corresponding values in a real three-dimensional (3D) device.
Yet, our modeling investigation should provide a deeper insight into the underlying physics of fluid flow and its related particle-fluid-wall interactions.

The trajectory of particles of various sizes obtained by using our developed algorithm is shown in Fig.~\ref{fig_res_pinched} for the case that width of particle-containing solution within the pinched segment is $w^\text{pcs}_\text{pinch}\approx 6~\mu m$.
The results demonstrate a size-based separation of sufficiently large particles.
It can be discerned that small particles cover a relatively wide portion of the expanded region which can also overlap with particles of other sizes, \textit{e.g.}, particles of diameter $1~\mu m$ and $5~\mu m$,
whereas sufficiently large particles end up moving on an almost identical particle size-dependent trajectory within the expanded region.

This behavior can be explained by considering the fact that particles become confined to a limited space in proximity of a sidewall within the pinched segment.
The center of particles with radii larger than $w^\text{pcs}_\text{pinch}$ falls outside of particle-containing stream within the pinched segment.
As a result of being pressed against a sidewall of pinched segment, these particles are sorted along the sidewall following distinct particle size-dependent streamlines, which diverge within the expanded region enabling high-resolution separation of particles.

In contrast, the center of particles with radii smaller than $w^\text{pcs}_\text{pinch}$ falls inside particle-containing stream within the pinched segment, because of which they can occupy a region of width of 
$w^\text{pcs}_\text{pinch}-d/2$.
For a shallow channel, this relatively narrow strip is theoretically broadened by a factor approximately equal to expansion ratio, $\psi$, which gives the width of space within the expanded region occupied by particles of diameter $d$ equal to $w^{p}_\text{exp.}=\psi(w^\text{pcs}_\text{pinch}-d/2)$.
This width can be relatively small for particle sizes close to $w^\text{pcs}$.
For example, particles of diameter $d=10~\mu m$ occupy a space of $w^{p}_\text{exp.}=20~\mu m$ wide within the expanded region (only twice as large as particle diameter), while the corresponding width is 
$w^{p}_\text{exp.}=70~\mu m$ and 
$w^{p}_\text{exp.}=110~\mu m$ for particles of diameter $d=5~\mu m$ and $d=1~\mu m$, respectively.

\begin{figure*}[!bt]
	\centering
	\includegraphics[width=1\textwidth, cfbox=blue 2pt 0pt]{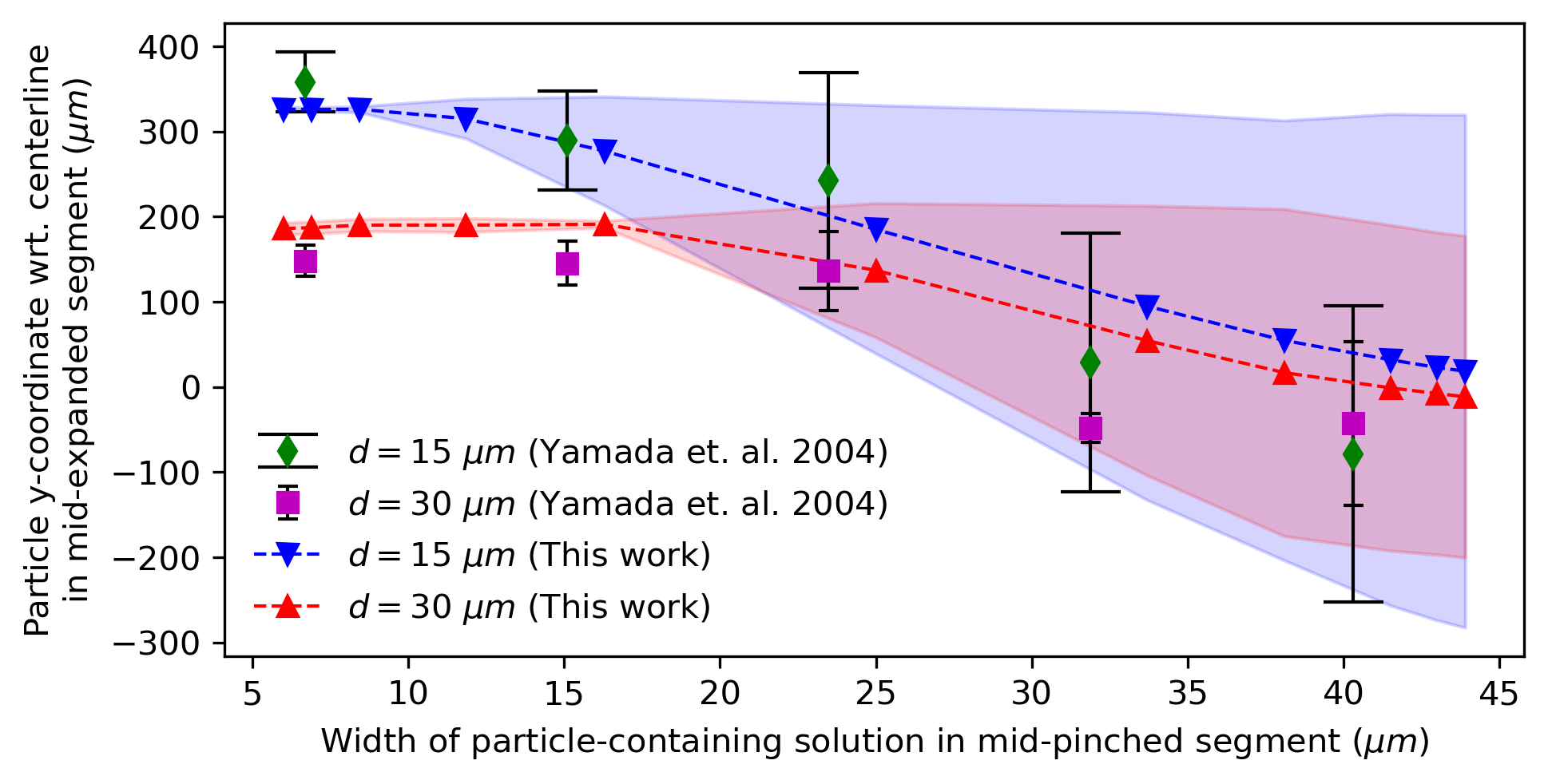}
	\caption{
		Variations of y-coordinate of particles of diameter $15~\mu m$ and $30~\mu m$ within mid-expanded segment of device as a function of width of particle-containing solution in mid-pinched segment 
		obtained by using Algorithm \ref{alg:pt} compared to experimental results reported by Yamada \textit{et. al.} \cite{yamada_pinched_2004}.
		Shaded areas show strips within expanded region, 
		determined from modeling,
		that can be occupied by particles of corresponding sizes.
		}
	\label{fig_res_pinch_comparison}
\end{figure*}

A quantitative comparison between 
our modeling results and those from Yamada \textit{et. al.} \cite{yamada_pinched_2004} 
has been made in Fig.~\ref{fig_res_pinch_comparison}
with regard to how particles occupy the expanded region.
Although a fairly good agreement is observed between the modeling and experimental results, there are discrepancies that might be attributed to the fact that our modeling is 2D while the real device in the work of Yamada \textit{et. al.} \cite{yamada_pinched_2004} was 3D. 
Numerical errors can be another source of such discrepancies that need to examined further in the future.

%% file: results_performance_evaluation.tex
\subsection{Performance evaluation}
\label{sec_res_performance_eval}
\begin{figure}[!bt]
	\centering
	\includegraphics[width=\textwidth, cfbox=blue 2pt 0pt]{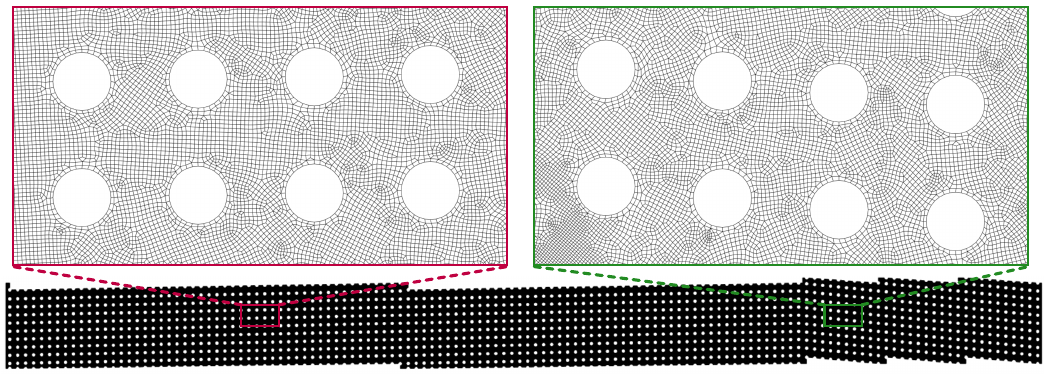}
	\caption{Model geometry and mesh used for evaluation of performance of the developed algorithm and its implantation code. The bounding box dimensions of the domain are about 3.64 mm $\times$ 0.315 mm.
	}
	\label{fig_res_performance_mesh}
\end{figure}

In this section, we present the results obtained from evaluating the performance of our developed algorithm and its implementation in C++ language.
Unless otherwise stated, we have used a computer with 
16.0 GB RAM and a 
13th Gen Intel\textsuperscript\textregistered~Core\textsuperscript{TM} i9-13980HX Processor.
The model for a condenser-sorter DLD system together with the generated mesh used in this section are shown in Fig.~\ref{fig_res_performance_mesh}.
The hybrid mesh consists of 948,262 nodes and 934,956 elements.

\subsubsection{Locating hosting cell's computational cost}
\label{sec_res_performance_eval_locating_hosting_cell}
We used a streakline consisting of 9 particles of $1~\mu m$ in diameter injected from the device inlet and performed particle tracking simulation until the particles reached the device outlet.
We studied both the original (ASG) and modified (MASG) variants of auxiliary structure grid algorithm. 
For the current mesh, all hosting cells could be located without using additional layers of cells around the core cells associated with a \textit{closest node}.
However, in order to study the effects of larger spiral levels that would be needed for a low-quality mesh, we examined different spiral levels of $\Phi=1,~2,~\text{and}~3$.
In addition, in order to study the effects of the number of circumferential vertices that form a particle, we considered four different cases: $N=4,~8,~16,~\text{and}~32$.
Furthermore, in order to investigate the effects of auxiliary grid bin size, we experimented auxiliary grids with seven different sizes.
The finest auxiliary grid consists of bins of the same size as the largest face/edge found in the mesh storing the continuous phase results, \textit{i.e.}, $\sim 2.6~\mu m$, while the bins in the largest auxiliary grid are 12 times as large as this value.
The number of bins of different auxiliary girds is shown in Table~\ref{tab_aux_grids}.
\begin{table*}[!t]
	\centering
	\caption{Number of bins of different auxiliary grids reported in Fig.~\ref{fig_res_performance_asg_masg}.}
	\begin{tabular*}{\textwidth}{@{\extracolsep{\fill}}||l c c c c c c c||} 
		\hline
		Dimensionless bin size & 1 & 2 & 4 & 6 & 8 & 10 & 12 \\ [0.5ex] 
		Number of bins & 1379x119 & 689x59 & 344x29 & 229x19 & 172x14 & 137x11 & 114x9\\  [0.5ex] 
		\hline
	\end{tabular*}
	\label{tab_aux_grids}
\end{table*}

\begin{figure}[!bt]
	\centering
	\includegraphics[width=\textwidth, cfbox=blue 2pt 0pt]{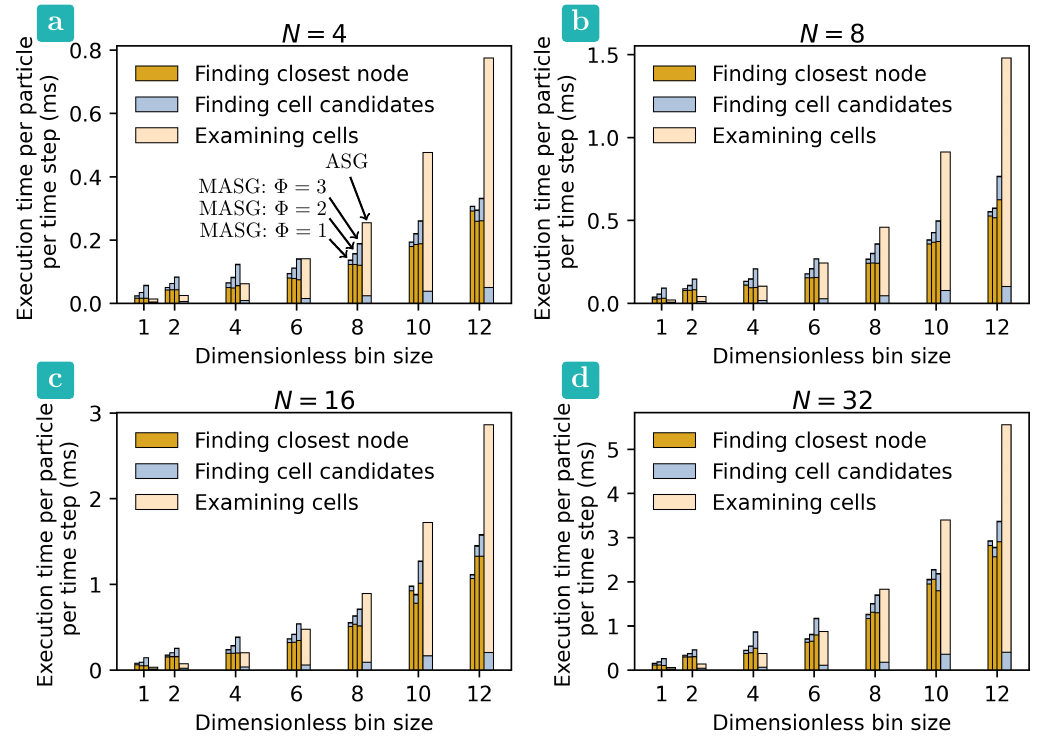}
	\caption{Execution time per particle per time step 
		as a function of bin size
		measured for different tasks included in the auxiliary structured grid (ASG) algorithm and its modified variant (MASG) by using various numbers of cell layers ($\Phi=1,~2,~\text{and}~3$) to be examined for locating hosting cells.
		The bin size of auxiliary grids is nondimensionalized by the length of largest edge ($\sim 2.6~\mu m$) within the main mesh storing the continuous phase results.
		The number of bins of different auxiliary grids can be found in Table~\ref{tab_aux_grids}.
	}
	\label{fig_res_performance_asg_masg}
\end{figure}

The obtained results are shown in Fig.~\ref{fig_res_performance_asg_masg}.
There are two tasks contributing in the computational cost of ASG algorithm: 1. finding the entire candidate cells from the corresponding bins of ASG, and 2. examining the candidate cells to determine the hosting cell. 
The computational cost of MASG algorithm can be fragmented into three parts: 1. finding the closest node, 2. finding the candidate cells sufficiently close to the closest node, and 3. examining the candidate cells to locate the hosting cell.
There are multiple important observations as discussed in the following.
\begin{itemize}
	\item The computational cost is proportional to the number of vertices considered on the perimeter of particles.
	\item In the case of original ASG, the computational cost is dominated by examining the candidate cells.
	\item  The computational cost of MASG is dominated mainly by finding the closest node, while the candidate cells examination is much less expensive (not visible in the chart due to much smaller execution times).
	\item The original ASG method is less expensive when using auxiliary grids with sufficiently small bins, while MASG is more efficient when using relatively coarser auxiliary grids.
\end{itemize}
Considering that the current results have been obtained by using a processor with only 8 performance threads, the modified variant of ASG method is expected to perform more efficiently by using the vectorized computation for finding closest node in conjunction with more advanced computational systems, in which case the modified ASG may become more economic than the original ASG technique regardless of the auxiliary grid size.
Exploring the full potential of this approach and determining its limit requires a more rigorous investigation and can be the subject of future studies.

\subsubsection{Parallel processing}
\label{sec_res_performance_parallel}
\begin{figure}[!bt]
	\centering
	\includegraphics[width=0.55\textwidth, cfbox=blue 2pt 0pt]{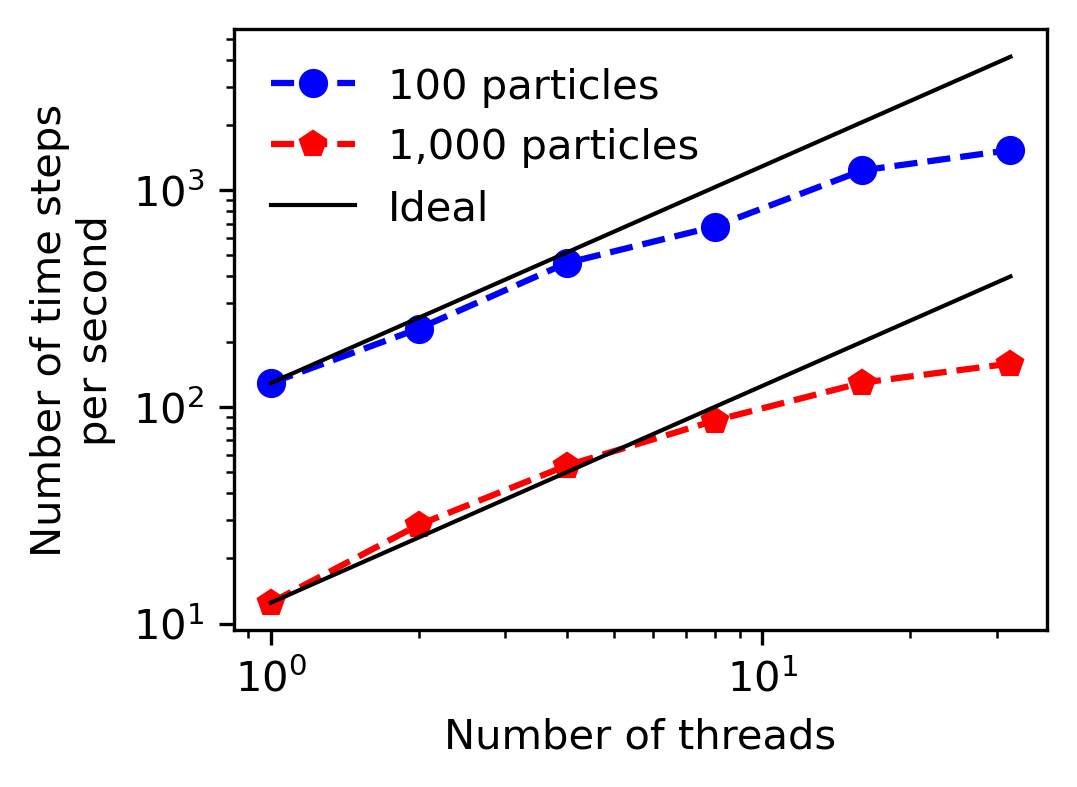}
	\caption{Parallel processing performance of the implementation code used for tracking of $1~\mu m$ particles in the condenser-sorter DLD system described in the text.
	}
	\label{fig_res_performance_parallel}
\end{figure}

We used OpenMP in our implementation code to enable parallel processing of particle tracking simulations by using multiple threads on a single machine.
We ran particle tracking simulations for 10,000 time steps by using the MASG algorithm with a spiral level of $\Phi=1$.
The obtained results are shown in Fig.~\ref{fig_res_performance_parallel}.
It can be discerned that the speedup increases almost linearly with the number of threads for up to 8 threads.
The speedup starts to become less significant for larger numbers of threads.
The main reason for deviation from the ideal speedup characteristics is that not all threads of the processor used in this work were the same.
The processor had 8 Performance-cores with a maximum Turbo Frequency of 5.60 GHz, as well as 16 Efficient-cores with a maximum Turbo Frequency of 4.00 GHz.
As a result, the simulations performed by using any subset of the 8 Performance threads are significantly more efficient than those leveraging a mixture of Performance and Efficient threads as shown in the last two data points, \textit{i.e.}, 16 and 32 threads.

%% file: conclusion.tex
\section{Conclusion}
\label{sec_conclusion}

In this work, we proposed a tracking algorithm for finite size particles.
While benefiting from the low computational cost of one-way coupling techniques to capture fluid-particle interactions, our method models interactions between particle and nearby solid objects explicitly by tracking circumferential points of particle efficiently and applying a boundary condition as needed to ensure impenetrability of solid objects.
In order to locate hosting cell efficiently, we proposed a modified ASG method.
Our search algorithm still benefits from auxiliary structured grid storing information of nodes/cells of main unstructured grid residing inside each structured cell.
However, instead of searching over the entire candidate cells within the corresponding bins of auxiliary grid, 
the closest node of unstructured grid to particle is found by processing the nodes within the corresponding bins in parallel, first, 
while an outward spiral search is performed over a smaller number of candidate cells in proximity of the pinpointed closest node afterwards to locate hosting cell of particle.
We investigated the computational cost of the original and the modified variants of ASG-based locating hosting cell algorithms.
We observed that the
the original ASG-based algorithm is faster when using sufficiently fine auxiliary grids of the order of the cell size found within the unstructured mesh, while the modified variant can outperform its original counterpart when using relatively coarser auxiliary grids.
We validated our algorithm experimentally by injecting $1.3~\mu m$ and $5~\mu m$ particles into a microfluidic DLD device 
and visualizing their motion using fluorescence microscopy. 
Our modeling work successfully demonstrated the zig-zag and bumping displacement modes that we observed in our experiments.
We also found a reasonable agreement between our modeling results and experimental data from the literature when determining critical diameter of multiple DLD structures.
In addition, we modeled a microfluidic device with pinched flow and validated our results against experimental data from the work reported by Yamada \textit{et. al.} \cite{yamada_pinched_2004}.
An acceptable agreement was found when comparing the particle trajectory results obtained from our modeling with those from experiments.
We also demonstrated that the algorithm can benefit from parallel processing on multiple threads by using OpenMP.
We observed a $\sim 8$x speedup by using 8 Performance threads.
The implementation of MPI can be considered in the future to further enhance the performance of the framework by using distributed systems.
The developed algorithm can find value in applications wherein 
particle-wall interactions play an important role, \textit{e.g.}, design and optimization of microfluidic devices for cell sorting, particle separation, exosome isolation, hydrodynamic filtration, etc.

%% file: main.bbl
\providecommand{\href}[2]{#2}\begingroup\raggedright\endgroup